\documentclass[final]{svjour3}
\usepackage{graphicx}
\usepackage{rotating}
\usepackage{amssymb}
\usepackage{siunitx}
\usepackage{array}
\newcolumntype{M}[1]{>{\centering\arraybackslash}m{#1}}
\DeclareSIUnit\sq{\ensuremath{\Box}}
\usepackage{mathptmx}
\usepackage[numbers,sort]{natbib}
\usepackage{xcolor}

\makeatletter

\journalname{Journal of Low Temperature Physics}

\bibpunct{[}{]}{,}{n}{}{,}

\begin{document}

\newcommand{\hdblarrow}{H\makebox[0.9ex][l]{$\downdownarrows$}-}
\title{A multi-chroic kinetic inductance detectors array using hierarchical phased array antenna}

\author{Shibo Shu$^{1,\dagger}$ \and Andrew Beyer$^{2}$  \and Peter Day$^2$ \and Fabien Defrance$^1$ \and Jack Sayers$^1$ \and Sunil Golwala$^1$}

\institute{$^1$ California Institute of Technology, Pasadena, California 91125, USA\\
$^2$ Jet Propulsion Laboratory, California Institute of Technology, Pasadena, California 91109, USA\\
$^\dagger$\email{shiboshu@caltech.edu}}

\maketitle

\begin{abstract}
We present a multi-chroic kinetic inductance detector (KID) pixel design integrated with a broadband hierarchical phased-array antenna. Each low-frequency pixel consists of four high-frequency pixels. Four passbands are designed from 125 to 365 GHz according to the atmospheric windows. The lumped element KIDs consist of 100-nm thick AlMn inductors and Nb parallel plate capacitors with hydrogenated amorphous Si dielectric. Two different coupling structures are designed to couple millimeter-wave from microstrip lines to KIDs. The KID designs are optimized for a 10-m-class telescope at a high, dry site, for example, the Leighton Chajnantor Telescope. Preliminary measurement results using Al KIDs are discussed.


\end{abstract}

\section{Introduction} 

Multi-chroic kinetic inductance detectors (KIDs) are an essential technology for future ground-based telescopes~\cite{Day:2003a, Perotto:2020a}. Mapping speed, one of the key factors describing total power instruments, is mainly determined by the number of pixels. In the mm/sub-mm wave ranges, the optimal pixel size and spacing vary with frequency. For a multi-chroic pixel, where more than two bands exist, 
not all bands can achieve the best spacing, due to their identical pixel spacing. One way to solve this problem is to use a hierarchical focal plane array~\cite{Cukierman:2018hierarchical}. Here we present a hierarchical pixel design using a slot-antenna array with KIDs. The long-term goal of this work is to make a hierarchical focal plane array covering 75-415 GHz with 6 bands~\cite{stacey2006:LWCam}, detailed in Table~\ref{tab:1}, for ground-based large telescopes, like the Leighton Chajnantor Telescope. As a technology demonstration, here we present a 2-scale design with 4 frequency bands from 125 to 365 GHz (Band 2-4) and preliminary measurement results.

\begin{table}
    \centering
    \begin{tabular}{c|c|c|c|c|c|c}
    \hline Band & 1 & 2 & 3 & 4 & 5 & 6 \\
    \hline$\lambda$ [mm] & 3.3 & 2 & 1.33 & 1.05 & 0.85 & 0.75 \\
    \hline$\nu$ [GHz] & 90 & 150 & 230 & 290 & 350 & 400 \\
    \hline$\Delta \nu$ [GHz] & 35 & 47 & 45 & 40 & 34 & 30 \\
    \hline $\mathrm{P}_{\text {opt}}$ [pW] & 4.0 & 8.6 & 8.5 & 9.9 & 12 & 17 \\
    \hline $l_{\text{antenna}}$ [mm] & 6.66 & 6.66 & 3.33 & 3.33 & 1.66 & 1.66 \\
    \hline Optimization results\\
    \hline$f_r [\mathrm{MHz}]$ & 125 & 175 & 225 & 275 & 325 & 375 \\
    \hline$A_{Ctot} [\mathrm{mm}^{2}]$ & 7.2 & 3.7 & 2.2 & 1.5 & 1.0 & 0.8\\
    \hline$Q_{i}$ [$\times 10^{4}$] & $3.4$ & $2.0$ & 2.0 & 1.6 & 1.4 & 1.0 \\
    \hline$Q_{r}$ [$\times 10^{4}$] & $1.7$ & $1.0$ & 1.0 & 0.8 & 0.7 & 0.5 \\
    \hline$n_{\mathrm{qp}}$ [$\mu \mathrm{m}^{-3}$] & 2619 & 3422 & 2889 & 3022 & 3216 & 3901\\
    \hline$\tau_{\mathrm{qp}}$ [$\mu \mathrm{s}$] & 37 & 28 & 33 & 32 & 30 & 25\\
    \hline $\mathrm{NEP}_{\mathrm{photon}}$ [$\mathrm{aW} /\sqrt{\mathrm{Hz}}$] & 34 & 64 & 70 & 85 & 109 & 157\\
    \hline $\mathrm{NEP}_{\mathrm{gr}}$ [$\mathrm{aW} /\sqrt{\mathrm{Hz}}$]     & 21 & 34 & 40 & 45 & 51  & 62\\
    \hline $\mathrm{NEP}_{\mathrm{a m p}}$ [$\mathrm{aW} /\sqrt{\mathrm{Hz}}$]  & 4  & 11 & 13 & 17 & 24  & 38\\
    \hline $\mathrm{NEP}_{\mathrm{T L S}}$ [$\mathrm{aW} /\sqrt{\mathrm{Hz}}$]  & 4  & 9  & 12 & 17 & 23  & 34\\
    \hline $\mathrm{NEP}_{\mathrm{tot}}$ [$\mathrm{aW} /\sqrt{\mathrm{Hz}}$]    & 41 & 74 & 83 & 99 & 125 & 176\\
    \hline
    \end{tabular}
    \caption{Specifications of optical bands~\cite{Ji:2015a} and typical design parameters after optimization. In this optimization, AlMn is used with $Tc$=1~K and the operating temperature is 0.1~K.}
    \label{tab:1}
\end{table}

\section{Hierarchical focal plane design}

\begin{figure}
    \centering
    \includegraphics[width=0.95\textwidth]{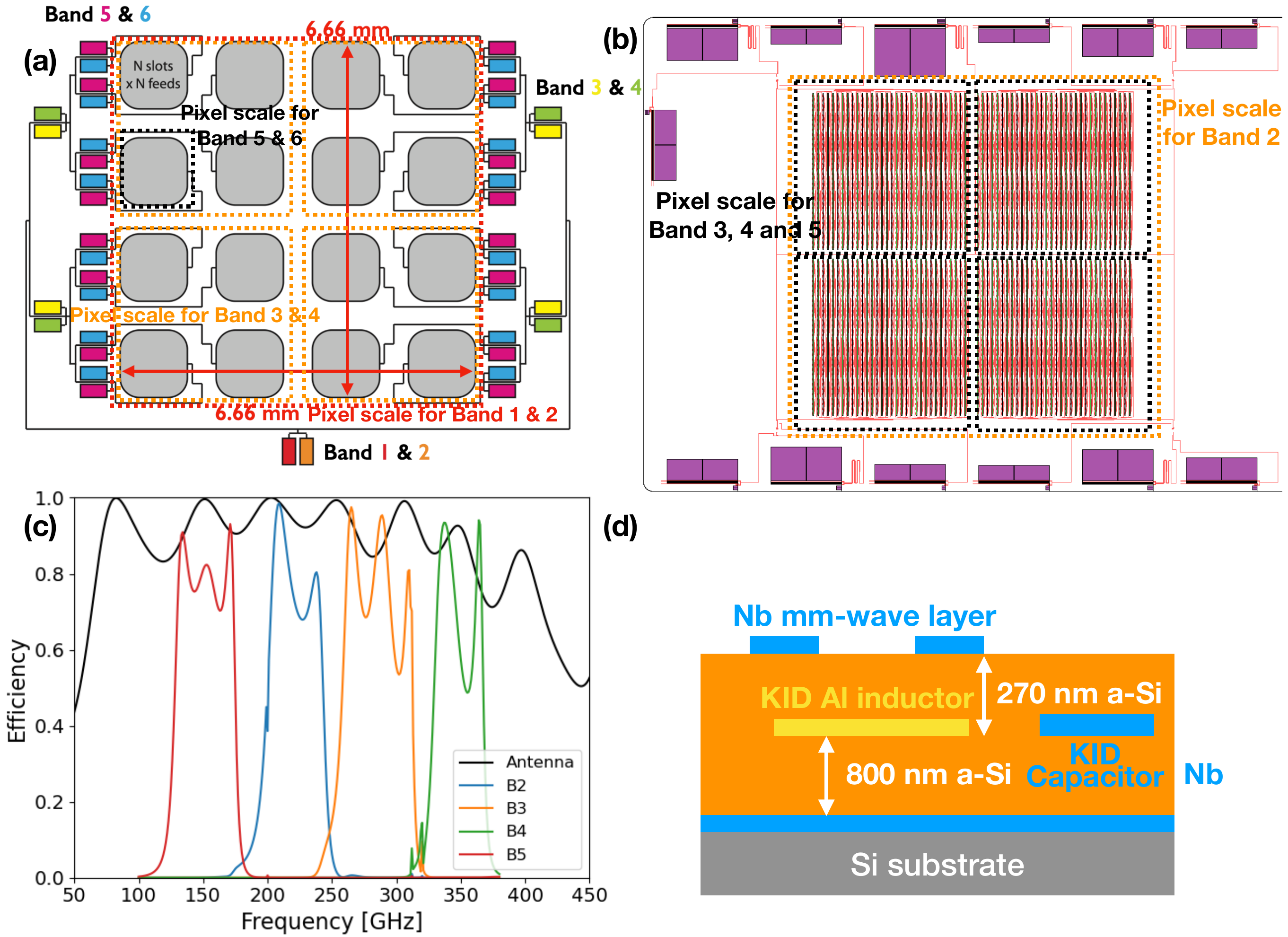}
    \caption{(a) The concept of the ultimate hierarchical pixel design with 6 bands. (b) The mask of a 2-scale pixel design including 13 KIDs. (c) Antenna efficiency assuming a perfect anti-reflection layer. Transmission of the BPFs network are also plotted. The total coupling is the product of the two results. (d) The material stack-up.}
    \label{fig:1}
\end{figure}

Resonant slot antenna array has been successfully used by the BICEP collaboration~\cite{Ade:2014bicep2design}. Compared to the resonant slot antenna, which has limited bandwidth, we follow the broadband non-resonant slot antenna design used in the MUSIC instrument~\cite{Goldin:2004a,Golwala:2012a}. This design covers 75-415~GHz allowing 6 bands to be placed inside, shown in Fig.~\ref{fig:1}c. 

To test the hierarchical concept, we made a 2-scale test pixel. The smallest antenna has a slot length of 3.33~mm for Band 3-5, and four small antennas are coherently added together to be a large pixel for Band 2 (Fig.~\ref{fig:1}b). The band separation is realized by 4 bandpass filters, grouped up as a network. The filter designs are modified from previous works~\cite{Kumar:2009a,Duan:2014a} and the performance of the network is shown in Fig.~\ref{fig:1}c.

\begin{figure}
    \centering
    \includegraphics[width=0.95\textwidth]{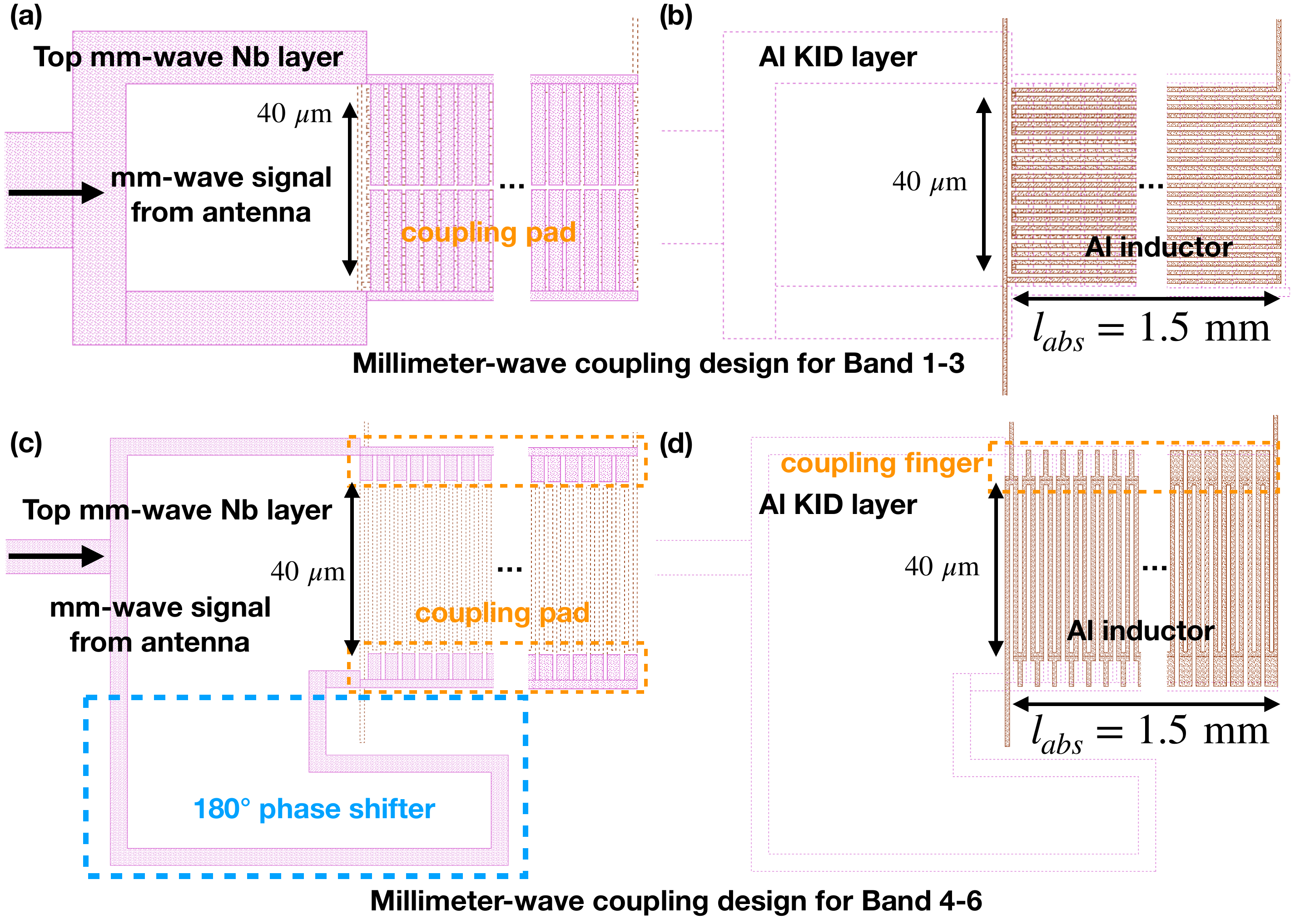}
    \caption{Two millimeter-wave coupler designs. (a) and (b) are the design for B1-3 and (c) and (d) are the design for B4-6.}
    \label{fig:2}
\end{figure}

\subsection{Millimeter-wave coupler}

To couple millimeter-wave signal to KIDs, we designed two capacitive couplers shown in Fig.~\ref{fig:2}. Both couplers have a power divider to separate the signal into two microstrip lines. Those coupling pads next to the microstrip lines decrease their characteristic impedance, so the microstrip lines are narrowed for impedance matching. The vertical size of the inductor is 40~$\mathrm{\mu m}$ and the horizontal size of the inductor is $l_{abs}$ used in optimization. $l_{abs}=1.5~\mathrm{mm}$ is used in the final design. Both couplers have the same inductor area and volume.

For low frequency bands (Band 1-3), the lossy AlMn inductor is designed to follow the signal traveling direction (Fig.~\ref{fig:2}b) as a lossy microstrip line and the coupling pads cover all the inductor area. For high frequency bands (Band 4-6), we let the inductor be perpendicular to the signal traveling direction (Fig.~\ref{fig:2}d) and coupling fingers are added on two sides of the inductor. A 180-degree phase shifter is used to alter the sign of the electromagnetic field, creating a voltage difference between the two coupling pads. The coupling strength is tuned by the area of the coupling fingers on the two sides of the inductor. With optimization, a near-uniform absorption is achieved in the inductor and the design is detailed in Ref.~\cite{Ji:2015a, Ji:2014design}.

\subsection{Optimization of the AlMn KIDs design}

The thickness of AlMn is designed to be 100~nm with a transition temperature of 1~K, a sheet resistance of \SI{0.069}{\ohm /\sq}, and a sheet inductance of \SI{0.11}{\pico\henry /\sq}. The \SI{1}{\um}-wide AlMn inductor has a magnetic inductance of \SI{6.85}{\nano\henry/\mm} and the kinetic inductance ratio is 26\%. Nb parallel plate capacitors with 800-nm thick aSi are used for tuning the resonance frequency. 

\begin{figure}
    \centering
    \includegraphics[width=0.95\textwidth]{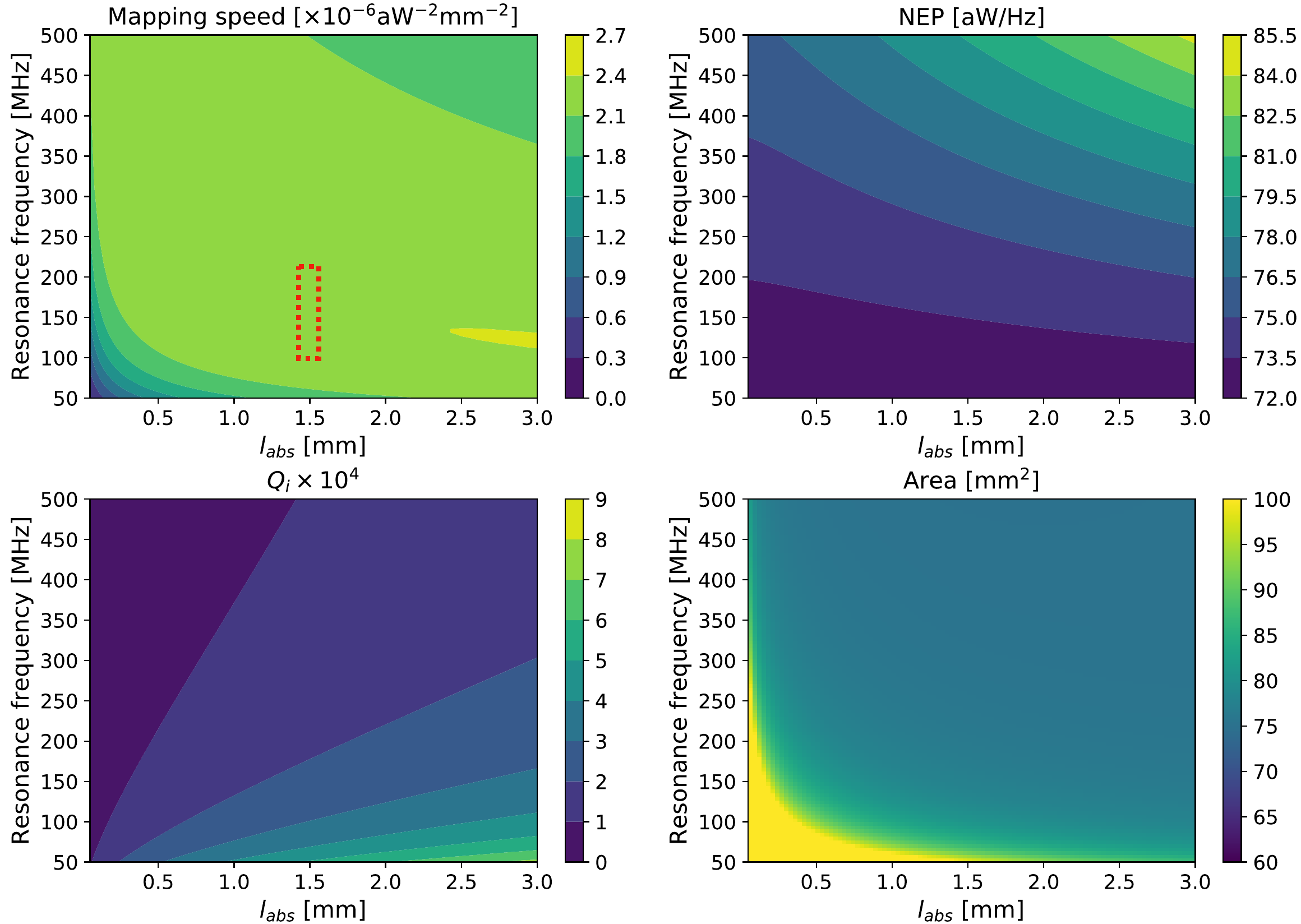}
    \caption{Calculations of 4 parameters in Band 2 KID optimization. The dashed red line shows where the final design is made. The optimal region is selected to have $Q_r>10^4$.}
    \label{fig:3}
\end{figure}

To find the optimal KID design, we selected the horizontal inductor size $l_{abs}$ (Fig.~\ref{fig:2}), and the resonance frequency $f_r$ as sweeping parameters. The coupling quality factor is set to be the same with the internal quality factor $Q_i$. The capacitor area is calculated from $f_r$. We first calculate the noise equivalent power (NEP), and then use the mapping speed $1/(\mathrm{NEP}^2\times \mathrm{Area})$, which takes the pixel area into account, to find the optimal design. $Q_r>10^4$ is set to have a reasonable readout multiplexing factor. The optimization result of Band 2 is shown in Fig.~\ref{fig:3}. Taking the optimization results of 6 bands, we decide to have $l_{abs}$=1.5~mm. The inductor volume is 3224~$\mu m^3$ for all KIDs. Resonance frequencies from B1 to B6 are designed from 100~MHz to 400~MHz with 50~MHz span for each band. Some typical optimization results are shown in Table.~\ref{tab:1}. In this test pixel, B2-5 KIDs are used. A microstrip line is used for readout.

\section{Dark and blackbody measurements of an Al test array}

\begin{figure}
    \centering
    \includegraphics[width=0.95\textwidth]{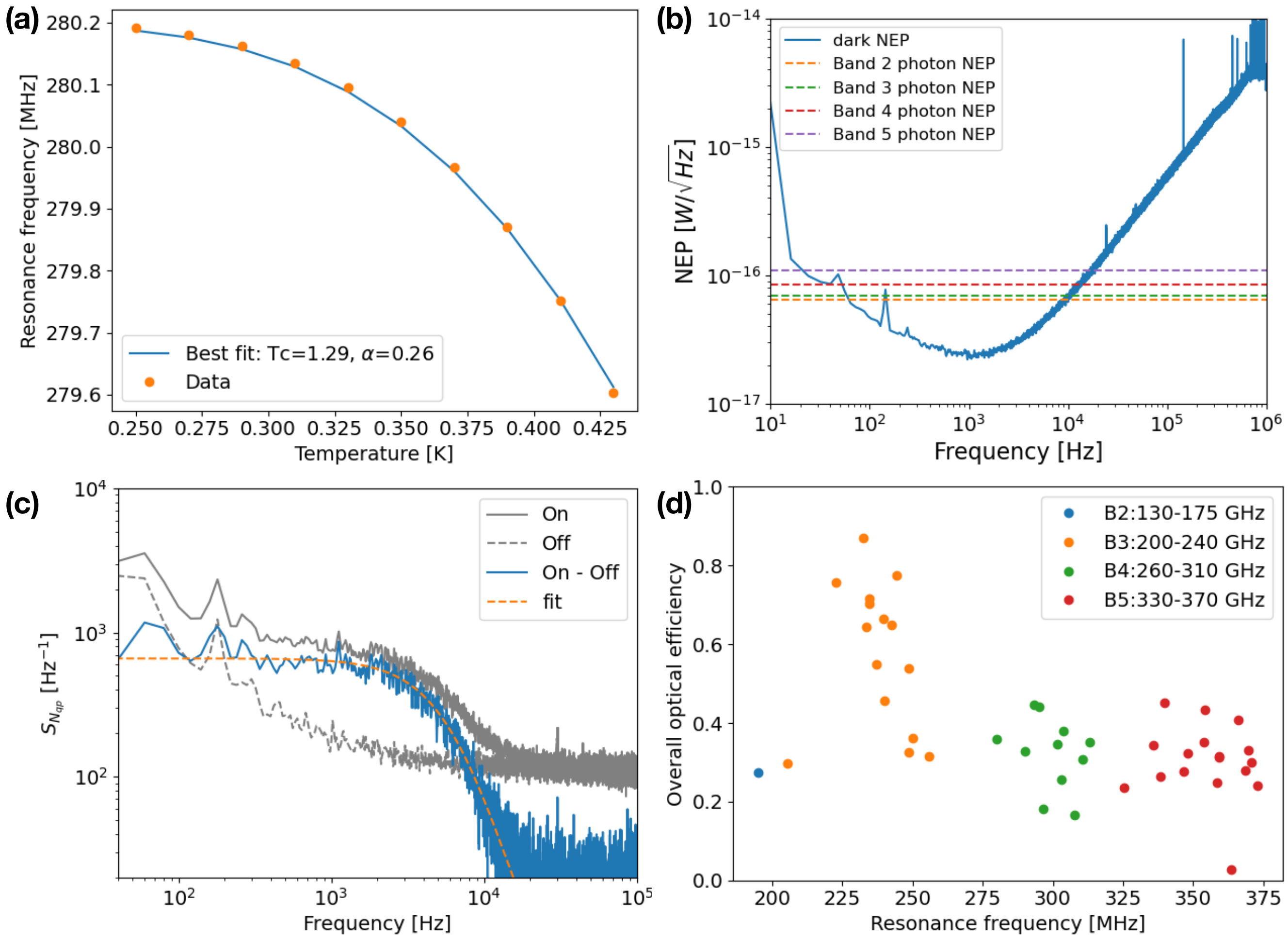}
    \caption{(a) Temperature sweep of resonance frequency. (b) Dark NEP in phase direction. (c) Quasiparticle generation-recombination noise at 290~mK in phase direction. (d) Optical efficiency (with pair breaking efficiency corrected) estimated from blackbody temperature sweep.}
    \label{fig:4}
\end{figure}

Al is used in our test device fabrication to verify the pixel concept. After fabrication, we performed one dark run and one blackbody run. 50 out of 56 KIDs are measured. By fitting the temperature sweep shown in Fig.~\ref{fig:4}, $T_c=1.29$~K and $\alpha=0.26$ are extracted, consistent with the values in design. The quasiparticle lifetime $\tau_{qp}=$\SI{100}{\micro\second}, shorter than expected, is fitted from cosmic-ray events at 247~mK. Using this data, the maximum lifetime $\tau_{max}=\SI{560}{\micro\second}$ is determined by including $\tau_{max}$ in the quasiparticle lifetime calculation~\cite{Zmuidzinas:2012a}. The dark NEP of \SI{2.2e-17}{\watt/\sqrt{\hertz}} is calculated from the measured noise spectrum $S_x$ using the following equation
\begin{equation}
\label{eqn:1}
    \mathrm{NEP} = \sqrt{S_{x}}\frac{dP}{dx}=\sqrt{S_{x}}  \frac{\Delta V}{\eta_{pb}\tau_{qp}} \frac{dn_{qp}}{dx},
\end{equation}
where $\Delta$ is the gap energy, $V$ is the inductor volume, and $\eta_{pb}$ is the pair breaking efficiency taken from Ref.~\cite{Guruswamy:2016a}. $dx/dn_{qp}$ is derived from the measured resonance frequency shifts and the quasiparticle density calculated from the simplified Mattis-Bardeen equation~\cite{Zmuidzinas:2012a}.

In the blackbody run, the array was measured at 290~mK by sweeping the blackbody temperature from 4~K to 6~K. The quasiparticle generation-recombination noise was measured in phase direction from a dark resonator, shown in Fig.~\ref{fig:4}c. At this temperature, the thermal $n_{qp}$ is very high and the possible quasiparticle density created by photons on this dark resonator is negligible. From fitting, the quasiparticle lifetime of \SI{25}{\micro\second} and quasiparticle density of \SI{2001}{/\cubic\um} are extracted, consistent with \SI{25}{\micro\second} and \SI{2026}{/\cubic\um} calculated using the simplified Mattis-Bardeen theory~\cite{Zmuidzinas:2012a}. 

Using the thermal $dn_{qp}/dx$, the optical efficiency is estimated for each resonator, shown in Fig.~\ref{fig:4}d. It has been shown that the responsivities derived from thermal and optical methods may give a difference up to a factor of 2~\cite{janssen2014equivalence}. In our high bath temperature case, the extra quasiparticles created by optical power is at the level of \SI{200}{/\cubic\um}, much smaller than the thermal one, so we suggest that this method can be used as an estimation of the optical efficiency. More rigorous results can be derived in a better experimental setup when the detectors are photon noise limited~\cite{Flanigan:2016b}. 

The Band 3 resonators show a higher efficiency than other bands, as the anti-reflection coating~\cite{Defrance:2018arcoating} is optimized for 170-300~GHz (-10~dB). In the next step, we will use Fourier transform spectrometer to detail the spectral characterization. 

\section{Conclusion}
We have shown a new multi-chroic kinetic inductance detectors array design using broadband hierarchical phased-array antenna. Preliminary results from dark and blackbody measurements are analyzed. The overall optical efficiency are estimated. Further measurements are needed for beam characterization.

\begin{acknowledgements}
This work has been partially funded by Department of Energy DE-SC0018126 and NASA 80NSSC18K0385.
\end{acknowledgements}

The datasets generated during and analysed during the current study are available from the corresponding author on reasonable request.

\bibliographystyle{JLTPv2}
\bibliography{Shu_lib.bib}
\end{document}